\documentstyle[referee]{mn}
\begin{document}
\title[Metallicity distribution of halo stars]{Metallicity distribution of halo
stars and minor merger processes of the Galaxy}
\author[Y. Lu, et al.]
{Y. Lu$^{1,2}$, K.S.Cheng$^{1}$, L.C. Deng$^{1,2}$\\
 $^1$Department of Physics, Honk Kong University, Hong Kong, China\\
 $^2$Beijing Astronomical Observatory, 20A Datun Road,
Chaoyang, Beijing 100012, China}
\maketitle
\markboth{ Y. Lu, K.S.Cheng\& L.C.Deng: Galactic halo stars and minor
merger of the galactic satellites}{}

\begin{abstract}
In this paper, a possible relation between the high dispersion in metallicity
of metal-poor halo stars and the minor merger processes in the history of the
Galaxy is presented. Observations show that satellite galaxies have been
falling into the gravitational well of the Galaxy and then disrupted by the
tidal
force through minor merger processes. As a result, the foreign populations
of stars in the satellites make considerable contributions to the Galactic
halo, therefore alter the intrinsic distribution of metallicities of the
halo stars. A model for the distribution of metallicities of halo stars with
$[Fe/H]<-2$ is made, which is constrained by observations.
We show that most of the metal-poor halo field stars come from the
satellite galaxies mergered into the Galaxy. Assuming the bulk of stars in a
satellite galaxy were formed in a cloud which had been enriched by
previous type II supernova events, our model reproduces the observed trends
in the metallicity distribution of the extremely metal-poor halo field
stars in the Galaxy. Taken all the parameters for our model, through mergering
satellite galaxies into the Galactic halo, a density of
$1.9\times 10^{-3}kpc^{-3}$ of extremely metal-poor halo stars in $-5<[Fe/H]<-4$
is predicted, which explains why no such stars have been observed so far.
\end{abstract}


\section{Introduction}
The origin of the Galactic stellar halo is a long-standing
problem. Theories of the formation of the Galaxy can generally be
viewed as variations on the monolithic collapse model of Eggen,
Lynde-Bell, \& Sandage (1962, hereafter ELS) and the chaotic
accretion model of Searle (1977) and Searle \& Zinn (1978,
hereafter SZ), or conbimations of the two models. The SZ picture
seems more popular in recent years( Wyse 1999 a,b; Gould et al.
1992; Irwin \& Hatzidimitrion 1995; Kuhn, Smith \& Hawley 1996).
SZ proposed a scenario in which the Galaxy was assembled through
the gradual merger of many sub-Galactic-sized clouds. Galactic
merger events generally can be divided into two classes:
(Kauffmann \& Charlot 1998) (a) minor mergers (i.e. accretion
events), which occur when one galaxy has less than a third of the
mass of the other one, and (b) major mergers, which happen when
the galaxies are within a factor of 3 in mass. For the goal of
the present work, only minor mergers are considered.

 Minor mergers involving satellite galaxies might have played a
 certain subtle role (Johnston, Hernquist \& Bolte 1996, hereafter JHB),
 although not being critical
for the formation and structure of the Galaxy. Observational
evidence associated with minor mergers, such as some quasar
activities(Bahcall, Kirhakos, \& Schneider 1995), Grand design
spiral structure, close companions (i.e. M51),  are shown.
Considerab observational evidance supporting accretion processes
has accumulated, eg. Through looking at the distribution of halo
stars in age and  metallicity, Unavane et al.(1996) concluded
that about 10\% of these stars could have been accreted from the
destruction of galaxies with stellar populations like those of
the dwarf spheroidal satellites of the Galaxy. Numerical
simulations also have confirmed the physical intuition that lumps
observed in the halo phase-space distribution could be associated
with accretion events (McGlynn 1990; Moore \& Davis 1994; Oh
etal.1995; Piatek \& Pryor 1995; Velazquez \& White 1995;
Johnston, Spergel \& Hernquist 1995; JHB; Kroupa 1997). The
observations of dwarf companions in the Galaxy, the discovery of
the Sagittarius dwarf galaxy (Ibata et al. 1994; Grillmair et al.
1995; JHB) have been observed. Furthermore, the possible
observable signatures left by that minor mergers is supported by
observations of satellites of the Galaxy that display evidence
for ongoing tidal interactions. In general, it can be concluded
through numerical simulation and observational data show that
satellite accretion is ongoing in the Galaxy and is likely to
have occurred often in the past.

To explore various formation scenarios of the Galactic halo,
Commonly employed tracers include the kinematics of halo field
stars as a function of $[Fe/H]$; the age distribution of Galactic
globular clusters; trends in cluster age with $[Fe/H]$ and
Galactocentric radius or height above the disk, and the
persistence of a cloud, i.e. the thin disk (Larson 1990; Majewski
1993). Low mass extremely metal-poor halo field stars have
lifetimes that are much greater than the age of the Galaxy so
that they will not evolve away from the main sequence, and due to
very long dynamical time scale of their orbital motion, the halo
stars do not dissipate their orbital energy. Hence, ``fossil"
information about the chemical composition patterns of the halo
is kept in these stars. In this paper, we consider the feasibility
of probing ``fossil" signatures of the formation of the Galaxy
through the frequency distribution of extremely metal-poor halo
field stars, as a function of $[Fe/H]$.

To address the properties of the extremely metal-poor halo stars
that appear in the range $[Fe/H]=-4$ to $-2.4$, many theoretical
models have been invoked in the framework of ELS prescription
(Cayrel 1986; Yoshii et al. 1995; Shigeyama \& Tsujimoto 1998;
Tsujimoto \& Shigeyama 1998; Tsujimoto et al. 1999; Lu et al.
2001). A challenging view against above models has been
suggested, in which, a Galactic halo is formed through the
disruption of many SZ fragments (Bekki 1998; Bullock et al. 2000,
hereafter BKW; Gilmore 2000; Gilmore \& Wyse 1998). The later
view argues that a diffuse stellar component produced by a large
number of tidally disrupted satellite galaxies, is perhaps
sufficient to account for most of the Galactic stellar halo.
Smaller galaxies collapse earlier when the density of the
universe was higher could be expected from the hierarchical
(Kravtsov et al. 1998; Kormendy \& Freeman 1998). Thus it is
likely that the satellite galaxies should have formed prior to
the epoch of main body of the Galaxy was assembled and would be
the building blocks of larger galaxies (Gilmore 2000).
Observationally, it is difficult to distinguish the foreign
contribution by the disrupted population from the normal stellar
halo component.

Motivated by above investigations, we present a model to
reproduce the observed metallicity distribution in the low
metallicity range ([Fe/H]$<-2.5$) of the Galactic halo stars by
taking into account the minor mergers of satellites. The
description and quantification of the model are given in Sect.2.
The results of our model and the conclusion are given in Sect.3.
\section{Model Description}
Up to a decade ago, searches for the first generation stars with strictly
the chemical composition left by Big Bang Neocleosynthesis (BBN)
had led to the result that
the observation limit towards the lowest metallicities (Beers Preston \&
Shectman.
1992, hereafter BPS) is now about $[Fe/H]=-4$. More than 100 stars with
metallicities
between $[Fe/H]=-4$ to $[Fe/H]=-3$ were found, while no stars at all with
$[Fe/H]=-5$ were discovered. These very metal-poor halo stars show a great
diversity in their
elemental abundances and therefore a scatter in their element-to-iron ratios
$[El/Fe]$ of order 1 dex. This scatter gradually decreases with increasing
metallicity and eventaully becomes the same as that of the mean metallicity.
\subsection{The basic assumptions}
To facilitate our model, the following working assumptions are adopted:
\begin{enumerate}
\item Our initial conditions assume that a Galactic halo interstellar medium
(ISM) consisting of a
homogeneously distributed single gas phase with primordial abundance
and a total mass of about $10^8M_\odot$.

\item Most of the metal-poor Galactic
stellar halo stars with a high dispersion in metallicity come from
accretion of satellites which are tidally disrupted during
the merger processes within Galactic halo. If such tidal components
were to maintain spatial and kinematic coherence in the lifetime of the
Galaxy, then a halo formed through the disruption of many different
satellite galaxies (SZ fragments) would exhibit a diversity in its
phase-space, unlike a unique origin coming from a smooth, monolithic
collapse (ELS models) which ought to be featureless (JHB).
Our present model will use this distinguishing
feature of the tidal components to account for the high dispersion in
metallicity of the Galactic extremely metal-poor halo stars.

\item The stellar contents of a accreted satellite galaxy is approximated
with a single stellar population (SSP), which is defined as a group of stars
born at the same time in a chemically homogeneous cloud with a given
metallicity. The satellite galaxies mergered into the Galaxy at different
time have distinct properties and are assumed to be represented by SSP models
of various metallicities and ages. When mergered into the Galactic halo,
stars in the satellite with the age and metallicity of the SSP are simply
added to the halo and will make its contribution to the distribution
of metallicity of the halo. In fact, the star formation history of each satellites
shows a remarkable complication, it is very difficult to describe the formation
and evolution of satellites by using SSP model. We will discuss this in furture work.

\item The metallicities of SSPs goes
from $[Fe/H]=-4$ to approximately $-2.5$, as noted by McWilliam et
al. (1995). We limit the metallicity range of SSPs based on the long-lived
halo star lifetime. It is further assumed that the age of SSP
in the satellite galaxy is as long as that of the Galactic halo star.  An
approximation to the metallicity dependent mass-lifetime relation of the
Geneva Stellar Evolution and Nucleosynthesis Group (Schaller et al. 1992;
Charbonnel et al. 1993) is used to determine
the lifetime of SSP,
\begin{eqnarray}
\lefteqn {\log T=(3.79+0.24Z)-(3.10+0.35Z)\log M}\nonumber \\
 & &\mbox{} +(0.74+0.11Z)\log^2 M,
\end{eqnarray}
where T is the lifetime in $10^6$ yr, Z the metallicity in
solar metallicity $Z_\odot$ and M the mass in solar mass
$M_\odot$. Usually, theoretical
metallicity distribution of stars are constructed as a function of
metallicity $Z=\log(n_Z/n_H)_*-\log(n_Z/n_H)_\odot$, where
$\log \left(\frac{n_{Fe}}{n_{H}}\right)_*$ is the stars iron abundance and
$\log \left(\frac{n_{Fe}}{n_{H}}\right)_\odot$
is the solar iron abundance. Whereas
observations of stellar abundances are usually expressed in terms
of $[Fe/H]$, since the abundance of iron is the most easily
measurable. For halo stars, $[Fe/H]\neq Z$, because not all
elements are deficient by the same factor. The chemical evolution
models adopted in our paper is parameterized by the ``effective
yield", which is a measure of the efficiency of the enriching
processes. The yield determined from observed $[Fe/H]$ values is not
the effective yield of metals, but rather that of iron. Following
the argument of Lambert (1989) and Kurucz (1979), we also assumed that the
effective yield of metals will be 0.35 dex higher, so the relation between
metallicity $[Fe/H]$ and Z is,
\begin{equation}
[Fe/H]=Z-0.35.
\end{equation}
Given the final metallicity $[Fe/H]$ of a SSP, with low mass limit of $0.8\le
M\le 1$, we can solve Eqs.(1) to (2) to obtain the lifetime of SSP.
\end{enumerate}
\subsection{Quantifying the model}
The model can be quantified based on the assumptions discussed above. Two
conceptions are introduced: (a) the accretion mass $M_{acc}$ of halo
field stars which is defined as the total mass of the accreted halo field stars
by $N_{sat}$ isolated satellites tidally disrupted at time $\tau$, where
$N_{sat}$ is defined as the accreted number of satellites that occurred
during the elapsed time $\tau$, and $\tau$ is given by
\begin{eqnarray}
\tau&=&{2\over{3H_0}}[1-{1\over{(1+z)^{3/2}}}]~~,\nonumber
\end{eqnarray}
where $z$ is the redshift, $H_0$ is the Hubble constant.
(b) the total iron yeild $M_{Fe}$ of the accreted satellite
integrated along the isochrone with a given initial mass function (IMF).

We use the approximate analytic model of BKW which provides the accretion
histories of an ensemble of 100 Galactic-type
galaxies ($v_{circ}=200 km s^{-1}$, $v_{circ}$ is circular velocities of
galaxy). The model assumes a $\Lambda$CDM
cosmology with $\Omega_m=0.3, \Omega_\Lambda=0.7, h=0.7,$ and
$\sigma_8=1.0$, and provides masses, approximate disruption times, and
orbital evolution for each disrupted satellite,
where $h$ is the hubble constant in units of 100$kms^{-1} Mpc^{-1}$,
$\sigma_8$ is the rms fluctuation on the scale of $8h^{-1}Mpc$.
The total stellar masses of the disrupted satellites is estimated
by applying the
same hypothesis used by BKW: low mass satellites with virial
temperatures below $\sim 10^{4}K$ (while the circular velocity is
$v_{circ}\sim 30 km s^{-1}$) can only
accrete gas before the universe was reionized at $z=z_{re}$ ($z_{re}$ is
reionization redshift). If the parameters of $z_{re}=8$ and $f=0.3$, where
$f$ the reionization stellar mass at $z_{re}$, which is constrained so that
 the observable halos
have mass-to-light ratios in the range of observed dwarf satellites (a
range $f \sim 0.1 - 0.8$ is plausible), the total stellar mass of the
disrupted component at time $\tau$ is roughly $M_{*tid}=5\times
10^8h^{-1}M_\odot$.

Zhao et al. (1999, hereafter ZJHS) analyzed a tidal components (stars or
gas clouds) that are turned from a satellite galaxy into the Galactic halo.
They suggested that
the number of field halo stars which share the same proper motion with a
tidal disrupted stellar components is
\begin{eqnarray}
N_{f}=N_{HB}{{\sigma_{tid}}\over {\sigma_{halo}}}\le 70 ,
\end{eqnarray}
where $\sigma_{halo}\sim 3000\mu as yr^{-1}$ and $\sigma_{tid}\sim 100\mu
as yr^{-1}$ are the dispersion velocities of halo stars and tidal
disrupted stellar components, respectively. $N_{HB}=(6\pm 2)\times 10^{4}$
is horizontal (HB) stars (Kinman 1994). If a disrupted satellite galaxy
had the same stellar content $N_{halo}\sim 6\times 10^4$ as the surviving
halos (BKW), with $M_{*tid}$ and Eq.(3),
we could obtain the total disrupted halo field star's mass at time $\tau$
\begin{eqnarray}
M_{f}&=&{{N_{f}}\over{N_{halo}}}M_{*tid}~~,\nonumber\\
         &=&4.08\times 10^5 ({N_{halo}\over {6 \times 10^4}})^{-1}({h\over
{0.7}})^{-1}({N_{f}\over {70}})M_\odot~~,
\end{eqnarray}
The relation among $M_{f}$, $M_{acc}$
and $N_{sat}$ is given as
\begin{eqnarray}
M_{acc}=M_{f}N_{sat} ~~.
\end{eqnarray}
Since the accreted mass $M_{acc}$ is directly proportional to the number of
accreted satellites, it can become larger than the total Galactic halo mass
$M_{tot}$. Furthermore, the
accretion efficiency per unit time can be defined as the
ratio $f_{acc}=M_{acc}/M_{tot}$, which depends only on $N_{sat}$ for fixed
$M_{f}$ and $M_{tot}$.
When $f_{acc}=1$, the total mass of the stars with chemical enrichment is
the same as the primodial halo mass of the Galaxy.
Since the stars coming from the accreted satellites have been mixed into
the Galactic halo, the ratio
$M_{f}/M_{tot}$ determines the mixing efficiency in our model,
therefore gives the number of the minor mergers of satellites in the unit
volume that is needed to reach a certain value of
$f_{acc}$. Given $f_{acc}$, $M_{tot}$ and the mean integrated iron
yield $<M_{Fe}>$ for a typical disrupted satellite, the mean metallicity
of the halo stars could be determined.

Despite significant effort, theoretical prediction of the abundance and
properties of the satellites are far from being complete (Klypin et al.
1999; Kauffman et al. 1993). The
self-enrichment mechanism of galactic halo globular clusters
(Parmentier et al. 1999) is adopted here to model the
abundances of disrupted satellite galaxies. The basic idea is that the
cold
and dense clouds embedded in the hot protogalactic medium are assumed to
be the progenitors of satellite galaxies. A first generation of
metal-free massive stars form in the center regions of proto-galaxies. The
corresponding massive stars evolve very quickly, end their lives as Type
II supernova
(hereafter SNII) and eject $\alpha$, $r$-process and possibly a small
amount of light s-process elements into the ISM. A second generation of
stars is born in these compressed and enriched layers of ISM. These SSPs
can
recollapse and form a satellite galaxy.
 The mass
$M_{Fe}$ ejected in the ISM by a SNII whose progenitor mass $m$
is approximately given by Woosley \& Weaver (1995), which ranges from
$12M_\odot$ to $60M_\odot$. If
the mass distribution of the first generation of stars obeys the
universal Salpeter form (Salpeter 1955)
\begin{eqnarray}
&& dN=Cm^{-2.35}dm,\\
&& \int^{M_u}_{M_{l1}}Cm^{-2.35}dm=1 \nonumber,
\end{eqnarray}
where $dN$ is the number of stars with masses between $m$ and $m+dm$,
$M_u=60M_\odot$ is the upper mass limit, $M_{l1}=0.1M_\odot$ is the lower
mass limit for the IMF. Then integrated iron yield $M_{Fe}$ in the
accreted satellite galaxy is
\begin{eqnarray}
<M_{Fe}>=\int^{m_u}_{m_{l2}}Cm^{-2.35} M_{Fe}dm,
\end{eqnarray}
where $M_{l2}$ is the lowest star mass for a SNII event. The
mean metallicity of the halo field stars can be given by,
\begin{eqnarray}
[Fe/H]&=&\log{{N_{sat}<M_{Fe}>}\over {M_{tot}}}-\log({{n_{Fe}}\over
{n_H}})_\odot~~,\nonumber\\
 &=&\log{{f_{acc}<M_{Fe}>}\over {M_{f}}}-\log({{n_{Fe}}\over
{n_H}})_\odot~~,
\end{eqnarray}
where $\log({{n_{Fe}}\over{n_H}})_\odot$ is the solar iron abundance.
Because the minor merger process simply adds the stars of the disrupted
galaxy into the Galactic halo, the evolution of the abundance ratios as a
function of $[Fe/H]$ is almost
independent of the star formation timescale in our model.

Table 1 shows the accretion factor in unit volume needed to reach the mean
metallicities of the extremely metal-poor Galactic halo stars:
$[Fe/H]=-5.0$,$ -4.5$, $-4.0$, $-3.5$,$ -3.0$,$ -2.5$, $-2.0$. Also shown
in the table are the corresponding accreted satellite frequency $N_{sat}$.

Assuming that the disrupted satellites have the same
star counts as those of the surviving  satellites (BKW), we can deduce the
number of
metal-poor halo field stars $N$,  that is expected in
different metallicity bins,
\begin{eqnarray}
N &=&N_{field}M_{acc}/M_{*tid}~~,\nonumber\\
 &=&N_{field}N_{sat}M_{f}/M_{*tid}
\end{eqnarray}
The number of metal-poor stars predicted by our model binned in
1.0 dex grid are listed in Table 2.
For comparison, the observed data from a homogeneous intermediate
resolution sample of Ryan \& Norris (1991) also lists in Table 2.
Assuming that total halo stars distributed in a volume of
$\frac{4\pi}{3}R_G^3$ ($R_G=20kpc$), from Table 1 and Table 2, we find that
178 mergers are need to produce a density of $1.9\times 10^{-3}kpc^{-3}$ for
the metal-poor stars in the bin $-5<[Fe/H]<-4$.
\section{Conclusion and discussion}
We have developed a model for the early chemical enrichment of the
Galactic
halo stars. The aim of the model is to understand the frequency
distribution
and the scatter in the $[El/Fe]$ ratios of the observed extremely
metal-poor halo stars.

As can be seen in the Table 2, we have deduced the expected number
of extremely matel-poor stars (with $[Fe/H]<-4$) which should have been
observed. we expect about 14 model stars with
$-5<[Fe/H]<-4$ while the observation sample contains none. If the ratio of
stars in these two metallicity bins for this admittedly inhomogeneous
sample is representative for the Galactic halo stars, this would suggest a
genuine shortage of the most metal-poor stars. It is possible that
Population III stars have caused a pre-enrichment in the
satellites and the true population III stars already have disappeared
before the onset of satellites accreted by the Galaxy. Furthermore the
possiblity to see the true Galactic population III is much reduced by
accretion of those pre-enriched foreign stars.

The growing number of the halo stars coming from mergered satellites
can be characterized by the accretion efficiency $f_{acc}$, defined as the
ratio
of the mass $M_{acc}=N_{sat}M_{f}$ produced by the number $N_{sat}$
tidally accreted satellites and the Galactic halo mass $M_{tot}$ with
primordial abundance gas.
The enrichment history of the halo stars is mainly determined by the
mixing
efficiency which in turn is fixed by the ratio of the mass $M_{f}$ of
the accreted stars of the satellite disrupted event and
$M_{tot}$. The more mass $M_{f}$ is, the less satellite accretion
event is needed to reach a certain value of $f_{acc}$, therefore the mixing
is more efficient. $M_{field}$ also determines the average metallicity
$[Fe/H]$ of the halo stars for a given accretion efficiency. A Larger
$M_{field}$ leads to a lower mean halo stars metallicity and vice versa.

Our model shows that the Galactic halo stars are the outcome of
a large number of tidally disrupted satellite galaxies, which is in
agreement with the prediction of BKW. Our analysis supports
that the Galactic halo is formed through the disruption of many
SZ fragments.

The observed differences in element ratio patterns of extremely
metal-poor halo field stars is a naturally results of our model,
because these stars come from galaxies with different SSP.
However, the alpha elements, iron peak elements and heavy
elements, as a function of $[Fe/H]$ in satellite galaxies are
produced by SNeII alone. The abundance ratios of tidal disrupted
components are predicted to exhibit a large star to star scatter,
depending in detail on the abundance patterns of SN ejecta with
different progenitor masses (Tsujimoto \& Shigeyama 1998;
Shigeyama and Tsujimoto 1998).

Based on present evidence, we propose a possible model for the formation
of our Galactic halo in this paper. Do large galaxies form from
accumulation of many smaller systems which have already initiated star
formation? Does star formation begin in a gravitational potential well in
which much of gas is already accumulated? and how to distinguish a
disrupted population from a stellar halo formed by other means
observationally? Answers to such questions require complementary
observational approaches. Fortunately, One of the next two 'cornerstones'
of ESA's science programmes, Global Astrometric Interferometer for
Astrophysics (GAIA), will advance all these questions (Freeman 1993;
Gilmore 1999; ZJHS; Hernandez et al 2000; Perryam et al. 2001). GAIA's
main scientific goal is to clarify the origin and history of our Galaxy,
from a quantitative census of the stellar populations. It will advance
questions such as when the stars in our Galaxy formed, when and how it was
assembled. The complete satellite system was evaluated as part of a
detailed technology study (Perryam et al. 2001).

We are indebeted to Prof. Xia X.Y. for many fruitful discussions
concerning this paper. We also thank Prof. Zhao G., Dr. Zhang L.
and Dr.Dai, Z.G. for a useful discussions and friendly help to
our paper. This work is partially supported by a Croucher
Fundation Senior Research Fellowship, an outstanding Research
Award of the University of Hong Kong and the Ministry of Science
and Technology through grant G19990754.

\clearpage
\begin{table}
\caption{Accretion efficiency $f_{acc}$ and tidal disrupted number
$N_{sat}$ of
satellite galaxy.}
\begin{tabular}{lcc}
\hline
   $[Fe/H]$  & $f_{acc}$ & $N_{sat}$     \\
\hline \hline
$-5.0$ & $0.074h^{-1}$ & $1.26 10^1$
\\
 $-4.5$ & $0.233h^{-1}$ & $3.99 10^1$
\\
$-4.0$ & $0.737h^{-1}$ & $1.26 10^2$
\\
 $-3.5$ & $2.332h^{-1}$ & $3.99 10^2$
\\
 $-3.0$ & $7.374h^{-1}$ & $1.26 10^3$
\\
 $-2.5$ & $23.32h^{-1}$ & $3.99 10^3$
\\
 $-2.0$ & $73.744h^{-1}$ & $1.26 10^4$
\\ \hline
\end{tabular}
\end{table}

\begin{table}
\caption{Top: Relative frequency of stars in the homogeneous
intermediate resolution survy of Ryan \& Norris
(1991), and our model, binned with binsize 1.0 dex. Bottom: Absolution
numbers. the last row gives the number of stars per 1.0 bin which we
expect to be present, if our model gives a fair representation of the halo
metallicity distribution. The number of model stars is normalized to the
number of stars in the range $-4<[Fe/H]<-2.0$ in the survy sample of Ryan
\& Norris (1991). No stars was detected with confirmed $[Fe/H]<-4.0$, in
contrast to the predicted by the model}.
\begin{tabular}{lccc} \hline
$[Fe/H]$  & $[-3.0, -2.0]$ & $[-4.0, -3.0]$ & $[-5.0, -4.0]$     \\
\hline \hline
Ryan \& Norris & $0.943$ & $0.057$ & $0.00$
\\
 Model & $0.971$ & $0.097$ & $0.0096$
\\
\hline
 Ryan \& Norris & $100$ & $6.0$ & $0.0$
\\
 Expected & $1420$ & $142$ & $ 14$
\\ \hline
\end{tabular}
\end{table}

\end{document}